\documentclass[conference]{IEEEtran}
\IEEEoverridecommandlockouts

\usepackage{graphics}
\usepackage{hyperref}
\usepackage{amsmath}
\usepackage[pdftex]{graphicx}
\usepackage{multicol}
\usepackage{lipsum}
\usepackage{float}
\usepackage{lipsum, color}
\usepackage{mathtools, cuted}
\usepackage{cite}
\usepackage{amsmath,amssymb,amsfonts}
\usepackage{algorithmic}
\usepackage{graphicx}
\usepackage{textcomp}
\usepackage{xcolor}
\def\BibTeX{{\rm B\kern-.05em{\sc i\kern-.025em b}\kern-.08em
    T\kern-.1667em\lower.7ex\hbox{E}\kern-.125emX}}

\makeatletter
\newcommand*{\rom}[1]{\expandafter\@slowromancap\romannumeral #1@}
\makeatother
    
\begin{document}

\title{Inter-Numerology Interference Pre-Equalization for 5G Mixed-Numerology Communications}

\author{\IEEEauthorblockN{Buğra Alp~Çevikgibi\IEEEauthorrefmark{1},
                         Ali Murat~Demirtaş\IEEEauthorrefmark{1},
                         Tolga~Girici\IEEEauthorrefmark{1}, and
                        Hüseyin~Arslan\IEEEauthorrefmark{2}\IEEEauthorrefmark{3}}
                        
\IEEEauthorblockA{\IEEEauthorrefmark{1}Department of Electrical and Electronics Engineering,
TOBB University of Economics and Technology, Ankara, 06560, Turkey}
\IEEEauthorblockA{\IEEEauthorrefmark{2}Department of Electrical and Electronics Engineering, Istanbul Medipol University, Istanbul, 34810, Turkey}
\IEEEauthorblockA{\IEEEauthorrefmark{3}Department of Electrical and Electronics Engineering, University of South Florida, Tampa, FL, 33620, USA}
\textit{Email: bugra.cevikgibi@gmail.com, ademirtas@etu.edu.tr, tgirici@etu.edu.tr, huseyinarslan@medipol.edu.tr}
}

\maketitle

\begin{abstract}
This article proposes a pre-equalization method to remove inter-numerology interference (INI) that occurs in multi-numerology OFDM frame structures of fifth-generation New Radio (5G-NR) and beyond on the transmitter side. In the literature, guard bands, filters and interference cancellation methods are used to reduce the INI. In this work, we mathematically model how the INI is generated and how it can be removed completely for multi-numerology systems on the transmitter side.
\end{abstract}


\section{Introduction}
\IEEEPARstart{F}{ifth} generation (5G) wireless communication networks will require more flexible resource distribution to  support a variety of communication requirements. The use-cases of 5G Networks have been categorized into three service groups: enhanced mobile broadband (eMBB), massive machine-type communication (mMTC), and ultra-reliable and low latency communication (URRLC)\cite{NguyenHL20}. 

In the 5G physical layer, \emph{flexible numerology} is a key enabler of supporting diverse use cases. \emph{Flexible numerology} is directly related to subcarrier spacing (SCS) in Orthogonal Frequency Division Multiplexing (OFDM). SCS is closely related to the OFDM symbol duration, frequency selectivity of the channel fading, the channel coherence time, and also related to cyclic prefix (CP). Therefore, for different use cases, different numerologies can be suitable. While 4G systems support only one type of numerology with SCS of $15$ kHz, 5G supports multiple and flexible numerologies to satisfy diverse requirements of a flexible radio access technology \cite{ZaidiBTBSMKS16}.

Multiplexing different numerologies (i.e OFDM with different SCS) on the same frequency band cause in-band interference. This in-band interference is called inter-numerology interference (INI) and is caused by the out-of-band emissions which are emitted by numerologies.

Inserting a guard band between adjacent numerologies on the same frequency band is a conventional way to reduce the INI. However, inserting a guard band reduces the spectral efficiency of the system. Scheduling is also used to control INI and minimize the effect of INI \cite{DBLP:journals/access/YazarA18}. The interference between the numerologies depends on the transmitted numerologies' powers and subcarrier spacings. The works in  \cite{ChengZSR20}, \cite{ZhangZXMWX18}, \cite{MaoZXN20} analyze the INI that occurred between adjacent numerologies. In these studies, it is shown that INI is created on the transmitter side before transmission. In the paper \cite{MemisogluKBA20}, characteristics of the INI are exploited in order to insert subcarriers in some parts of the guard band. In \cite{ZhangZXMWX18}, an iterative interference cancellation algorithm is proposed on the receiver side. 

In this work, we analyze the INI by using discrete Fourier transform equations on frequency and time domain. Then, we model the INI on each transmitted subcarrier (i.e transmitted symbols) as weighted linear combinations of neighbor numerology's transmitted subcarriers. These weighted linear combinations are expressed as INI weight matrix. Our contribution is to almost eliminate the INI by multiplying the inverse of the W matrix as a pre-equalizer.

Simulation results reveal that INI is completely removed on transmitted subcarriers and theoretical minimum bit error rates (BER) for various modulations on a Rayleigh fading channel are achieved without using any guard band.

The rest of the paper is prepared as follows. In Section \rom{2}, the classical multi-numerology OFDM is revisited. In Section \rom{3}, the INI is analyzed by DFT and IDFT matrices, and INI on the numerologies are mathematically modeled. At the end of the Section \rom{3}, INI pre-equalization matrix is derived and illustrated. Section \rom{4} presents the simulation results of the proposed INI pre-equalizer. Finally, some concluding remarks and future works are provided in Section \rom{5}.

\section{System Model}

We consider a Downlink (DL) multi-numerology OFDM system. The standardized numerologies for 5G NR are \cite{ParkvallDFF17}

\begin{equation}
    \Delta f^{\mu } = 2^{\mu} \times 15 \mathrm{kHz}, \quad \mu = 0, 1, 2, 3, 4.
    \tag{1}
\end{equation}

The parameter $\Delta f^{\mu}$ is the subcarrier spacing of numerology $\mu$. Among these numerologies, $\mu=0, 1 , 2$ are allowed in the Frequency Range 1 (FR1), which is below $6$ GHz \cite{ParkvallDFF17}. The transmission spectrum is divided into two equal sub-bands, where each sub-band is assigned to one type of numerology. Assume two different numerologies, $\mu = 0, 1$ are multiplexed onto an OFDM frequency band\footnote{In the literature usually numerologies 0 and 1 are assumed. The same analysis can be easily carried out for numerology pairs $\mu = $ $0-2$ or $1-2$ or more than two different numerologies on the same frequency band.}. Without using any guard band, both numerologies can be generated as
\vspace{-0.2cm}
\begin{equation}
    {X}_{0}(k) = \sum _{k' = 0}^{N - 1}X_{0}(k')\delta (k - k')
    , \, X_{0}(k) = 0 \, \mathrm{for} \, \frac{N}{2} \leq k \leq N - 1
    \tag{2}
\end{equation}
\begin{equation}
    {X}_{1, q}(l) = \sum _{l' = 0}^{M - 1}X_{1, q}(l')\delta (l - l')
    , \, X_{1, q}(l) = 0 \, \mathrm{for} \, 0 \leq l \leq \frac{M}{2} - 1
    \tag{3}
\end{equation}
where $N$ and $M$ are the discrete Fourier transform and inverse discrete Fourier transform (DFT/IDFT) sizes of the numerologies. Complex modulation symbols are denoted as ${X}_{0}$ and ${X}_{1, q}$. $N = Q\times M$, where  $Q = 2$, $q = 0, 1$ and $\delta(.)$ is the  delta function. Index $q = 0, 1$ denotes time-multiplexed symbols of the numerology with wider subchannels (i.e. narrower symbol duration).
After subcarrier allocation into sub-bands, CP added discrete-time OFDM symbols for both numerologies that can be generated by the IDFT operation.
\begin{multline}
    y_{0}(n) = \frac{1}{N}\sum _{k = 0}^{N - 1}{X}_{0}(k)e^{\frac{j2\pi k \langle n - N_{cp}\rangle_{N}}{N}} 
    ,\\ 0 \leq n \leq N + N_{cp} - 1
    \tag{4}
\end{multline}
\begin{multline}
    y_{1, q}(m) = \frac{1}{M}\sum _{l = 0}^{M - 1}{X}_{1, q}(l)e^{\frac{j2\pi l \langle m - M_{cp} \rangle_{M}}{M}} 
    ,\\ 0 \leq m \leq M + M_{cp} - 1
    \tag{5}
\end{multline}
where $N_{cp}$ and $M_{cp}$ are CP duration of both numerologies and ${N_{cp} = Q \times M_{cp}}$. We denote modulo-$N$ operation as $\langle . \rangle_{N}$. To create the composite numerology OFDM signal, we must concatenate OFDM symbols $y_{1, q}(m)$  in the discrete-time domain.
\vspace{-0.2cm}
\begin{multline}
    y_{1}(n) = \sum_{m = 0}^{M + M_{cp} - 1}\sum_{q = 0}^{Q-1}y_{1, q}(m)\delta(n - m - q(M + M_{cp}))
    \tag{6}
\end{multline}

Composite numerology OFDM signal is generated as

\begin{equation}
    y_{composite}(n) = y_{0}(n) + y_{1}(n) 
    \tag{7}
\end{equation}

\section{Analysis of INI By DFT and IDFT Matrices}

$N_{cp}$-point CP-added $N$-point IDFT equation (4) can be written in a matrix form in equation (8) where $\bar{\omega}_{0}^{kn} = (e^{j2\pi/N})^{kn}$, which is the complex conjugate of ${\omega}_{0}^{kn}$.

$M_{cp}$-point CP-added $M$-point IDFT equation  (5) can be written in a matrix form as in equation (9), where $\bar{\omega}_{1}^{lm} = (e^{j2\pi/M})^{lm}$, which is the complex conjugate of ${\omega}_{1}^{lm}$.

$N$-point DFT Matrix for numerology 0 can be written as

\begin{equation}
\begin{bmatrix}
    X_{0}(0)     \\
    X_{0}(1)     \\
    \vdots       \\
    X_{0}(N - 1) \\
\end{bmatrix} \hspace{-0.1 cm}=\hspace{-0.1 cm} 
\begin{bmatrix}{\scriptscriptstyle}
   1      &\hspace{-0.1 cm} 1                    &\hspace{-0.3 cm}\cdots  &\hspace{-0.5 cm} 1                        \\      
   1      &\hspace{-0.1 cm} {\omega}_{0}         &\hspace{-0.3 cm}\cdots  &\hspace{-0.5 cm} {\omega}_{0}^{(N-1)}     \\
   \vdots &\hspace{-0.1 cm} \vdots               &\hspace{-0.3 cm}\ddots  &\hspace{-0.5 cm} \vdots                   \\
   1      &\hspace{-0.1 cm} {\omega}_{0}^{(N-1)} &\hspace{-0.3 cm}\cdots  &\hspace{-0.2 cm}{\omega}_{0}^{(N-1)(N-1)} \\
\end{bmatrix}
\hspace{-0.1 cm}
\begin{bmatrix}
    y_{0} (0)    \\
    y_{0} (1)    \\
    \vdots       \\
    y_{0} (N - 1)\\
\end{bmatrix} 
\tag{10}
\end{equation}

$M$-point DFT Matrix for numerology 1 can be written as

\begin{equation}
\begin{bmatrix}
    X_{1}(0)     \\
    X_{1}(1)     \\
    \vdots       \\
    X_{1}(M - 1) \\
\end{bmatrix} \hspace{-0.1 cm}=\hspace{-0.1 cm} 
\begin{bmatrix}{\scriptscriptstyle}
   1      &\hspace{-0.1 cm} 1                    &\hspace{-0.3 cm}\cdots &\hspace{-0.5 cm}1                         \\      
   1      &\hspace{-0.1 cm} {\omega}_{1}         &\hspace{-0.3 cm}\cdots &\hspace{-0.5 cm}{\omega}_{1}^{(M-1)}      \\
   \vdots &\hspace{-0.1 cm} \vdots               &\hspace{-0.3 cm}\ddots &\hspace{-0.5 cm}\vdots                    \\
   1      &\hspace{-0.1 cm} {\omega}_{1}^{(M-1)} &\hspace{-0.3 cm}\cdots &\hspace{-0.2 cm}{\omega}_{1}^{(M-1)(M-1)} \\
\end{bmatrix}
\hspace{-0.1 cm}
\begin{bmatrix}
    y_{1} (0)    \\
    y_{1} (1)    \\
    \vdots       \\
    y_{1} (M - 1)\\
\end{bmatrix}
\tag{11}
\end{equation}

In equations (8) and (9), first $N_{cp}$ and $M_{cp}$ rows are the CP's of $y_{0}$ and $y_{1}$, respectively. 

\vspace{-1.3cm}
\begin{strip}
\begin{gather}
\underbrace{
\begin{bmatrix}
    y_{0}(0)     \\
    y_{0}(1) \\
    \vdots                \\
    y_{0}(N_{cp} - 1)          \\
    y_{0}(N_{cp})              \\
    y_{0}(N_{cp} + 1)              \\
    \vdots                \\
    y_{0}(N + N_{cp} - 1)          \\
\end{bmatrix}
}_{y_{0}}= 
\frac{1}{N}
\begin{bmatrix}
    1      & \bar{\omega}_{0}^{(N-N_{cp})}   & \cdots & \bar{\omega}_{0}^{(N-1)(M - N_{cp})}      \\      
    1      & \bar{\omega}_{0}^{(N-N_{cp}+1)} & \cdots &  \bar{\omega}_{0}^{(N-1)(M - N_{cp + 1})}  \\
    \vdots & \vdots                          & \ddots & \vdots                                    \\
    1      & \bar{\omega}_{0}^{(N-1)}       & \cdots  & \bar{\omega}_{0}^{(N-1)(N-1)}            \\
    1      & 1                               & \cdots & 1                  \\
    1      & \bar{\omega}_{0}                & \cdots & \bar{\omega}_{0}^{(N-1)}                 \\
    \vdots & \vdots                          & \ddots & \vdots                                        \\
    1      & \bar{\omega}_{0}^{(N-1)}       & \cdots  & \bar{\omega}_{0}^{(N-1)(N-1)}            \\
\end{bmatrix}
\begin{bmatrix}
   X_{0}(0)
\\ X_{0}(1)
\\ \vdots 
\\ X_{0}(N - 1)
\end{bmatrix}
\tag{8}
\end{gather}
\vspace{-0.5cm}
\begin{gather}
\underbrace{
\begin{bmatrix}
    y_{1}(0)      \\
    y_{1}(1)  \\
    \vdots                 \\
    y_{1}(M_{cp} - 1)           \\
    y_{1}(M_{cp})               \\
    y_{1}(M_{cp} + 1)               \\
    \vdots                 \\
    y_{1}(M + M_{cp} - 1)           \\
\end{bmatrix}
}_{y_{1}}=  
\frac{1}{M}
\begin{bmatrix}
    1      & \bar{\omega}_{1}^{(M-M_{cp})}   & \cdots & \bar{\omega}_{1}^{(M-1)(M - M_{cp})}      \\      
    1      & \bar{\omega}_{1}^{(M-M_{cp}+1)} & \cdots &  \bar{\omega}_{1}^{(M-1)(M - M_{cp + 1})}  \\
    \vdots & \vdots                         & \ddots & \vdots                                    \\
    1      & \bar{\omega}_{1}^{(M-1)}       & \cdots  & \bar{\omega}_{1}^{(M-1)(M-1)}            \\
    1      & 1                               & \cdots & 1                  \\
    1      & \bar{\omega}_{1}                & \cdots & \bar{\omega}_{1}^{(M-1)}                 \\
    \vdots & \vdots                         & \ddots & \vdots                                    \\
    1      & \bar{\omega}_{1}^{(M-1)}       & \cdots  & \bar{\omega}_{1}^{(M-1)(M-1)}            \\
\end{bmatrix}
\begin{bmatrix}
   X_{1}(0)
\\ X_{1}(1)
\\ \vdots 
\\ X_{1}(M - 1)
\end{bmatrix}
\tag{9}
\end{gather}
\end{strip}

Let us choose $P$ subcarriers of numerology 0 and $K$ subcarriers of numerology 1 at the neighbor boundary of both numerologies. That is, we take the last $P$ subcarriers of numerology 0 from $\frac{N}{2}-P$ to $\frac{N}{2}-1$ subcarrier indexes, and first $K$ subcarriers of numerology 1 from $\frac{M}{2}$ to $\frac{M}{2}+K-1$ subcarrier indexes, from equations (2) and (3) to analyze and remove the INI on subcarriers to be transmitted.

\subsection{INI from numerology 0 to numerology 1-0 and 1-1}

\begin{figure}[ht!] 
 \centering
 \includegraphics[width=0.45\textwidth] {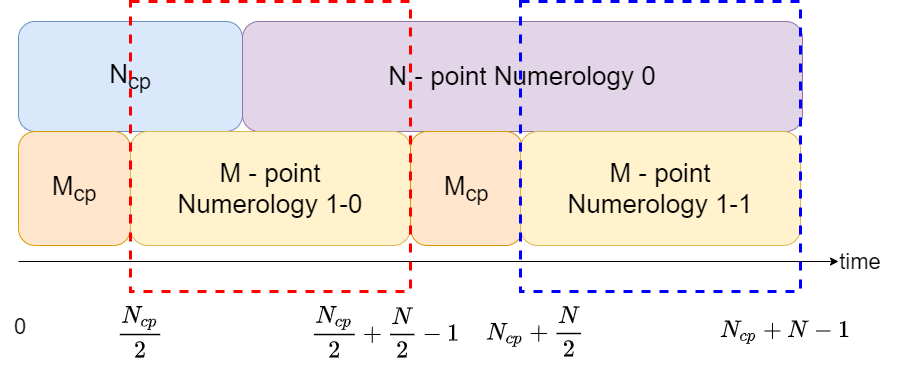}
 \caption{INI from numerology 0 to numerology 1-0 and 1-1 on time domain}
\end{figure}

The first part of numerology 0, shown in red dashed lines in Fig 1, between time samples $\frac{N_{cp}}{2}$ and $\frac{N_{cp}}{2} + \frac{N}{2} - 1$ of subcarriers from $\frac{N}{2}-P$ to $\frac{N}{2}-1$ causes INI on subcarriers from $\frac{M}{2}$ to $\frac{M}{2} + K - 1$ of numerology 1-0. This expression can be modeled as a matrix form in equation (12).

Eq. (12) can be simplified as in (13). Where, $\textbf{\emph{INI}}_{1,0}$ is a $K\times 1$-column vector that represents the INI on the first K subcarriers of numerology 1-0. $\textbf{\emph{X}}_{0}$ is the $P\times 1$-column vector of symbols to be transmitted of numerology 0 and $\textbf{\emph{W}}_{1,0}^{INI}$ is the $K\times P$ matrix describing how the INI is created.

\begin{equation}
    {\mathbf{INI}_{1,0}} = {\mathbf{W}_{1,0}^{INI}}  {\mathbf{X}_{0}}
\tag{13}
\end{equation}

The second part of numerology 0, shown in blue dashed lines in Fig 1, between time samples $N_{cp} + \frac{N}{2}$ and $N_{cp} + N - 1$ of subcarriers from $\frac{N}{2}-P$ to $\frac{N}{2}-1$ causes INI on subcarriers from $\frac{M}{2}$ to $\frac{M}{2} + K - 1$ of numerology 1-1. This expression can be modeled as a matrix form in equation (14), which also can be simplified as in  (15).

\begin{equation}
    {\mathbf{INI}_{1,1}} = {\mathbf{W}_{1,1}^{INI}}  {\mathbf{X}_{0}}
\tag{15}
\end{equation}

\vspace{-2.6 cm}
\begin{strip}
\begin{gather}
\underbrace{
\begin{bmatrix}
   INI_{1,0}\scriptstyle{(\frac{M}{2})}         \\
   INI_{1,0}\scriptstyle{(\frac{M}{2} + 1)}     \\
   \scriptstyle{\vdots}                         \\
   INI_{1,0}\scriptstyle{(\frac{M}{2} + K - 1)}  \\
\end{bmatrix} \hspace{-0.15 cm}=\hspace{-0.15 cm}
}_{\mathbf{INI}_{1, 0}}
\underbrace{
\begin{bmatrix}
   1         &\hspace{-0.25 cm}\cdots &\hspace{-0.25 cm}{\omega}_{1}^{(\frac{M}{2})(M - 1)}       \\      
   1         &\hspace{-0.25 cm}\cdots &\hspace{-0.25 cm}{\omega}_{1}^{(\frac{M}{2} + 1)(M - 1)}   \\
   \vdots    &\hspace{-0.25 cm}\ddots &\hspace{-0.25 cm}\vdots                            \\
   1         &\hspace{-0.25 cm}\cdots &\hspace{-0.25 cm}{\omega}_{1}^{(\frac{M}{2} + K - 1)(M-1)} \\
\end{bmatrix}
\hspace{-0.1 cm}
\frac{1}{N}
\hspace{-0.1 cm}
\begin{bmatrix}
   {\bar\omega}_{0}^{(\frac{N}{2}-P)(\frac{N_{cp}}{2})}     &\hspace{-0.4 cm}\cdots      &\hspace{-0.2 cm} {\bar\omega}_{0}^{(\frac{N}{2}-2)(\frac{N_{cp}}{2})}     &\hspace{-0.3 cm}{\bar\omega}_{0}^{(\frac{N}{2}-1)(\frac{N_{cp}}{2})}     \\      
   \vdots                                 &\hspace{-0.4 cm}\ddots      &\hspace{-0.2 cm} \vdots                                 &\hspace{-0.3 cm}\vdots                                 \\
   {\bar\omega}_{0}^{(\frac{N}{2}-P)(\frac{N_{cp}}{2} + \frac{N}{2} - 1)} &\hspace{-0.4 cm}\cdots      &\hspace{-0.2 cm} {\bar\omega}_{0}^{(\frac{N}{2}-2)(\frac{N_{cp}}{2} + \frac{N}{2} - 1)} &\hspace{-0.3 cm}{\bar\omega}_{0}^{(\frac{N}{2}-1)(\frac{N_{cp}}{2} + \frac{N}{2} - 1)} \\
\end{bmatrix}
}_{\mathbf{W}_{1,0}^{INI}}
\hspace{-0.2 cm}
\underbrace{
\begin{bmatrix}
   X_{0}\scriptstyle{(\frac{N}{2}-P)} \\
   \scriptstyle{\vdots}        \\
   X_{0}\scriptstyle{(\frac{N}{2}-2)}\\
   X_{0}\scriptstyle{(\frac{N}{2}-1)} \\
\end{bmatrix}
}_{\mathbf{X}_{0}}
\tag{12}
\end{gather}
\vspace{-0.5cm}
\begin{gather}
\underbrace{
\begin{bmatrix}
   INI_{1,1}\scriptstyle{(\frac{M}{2})}        \\
   INI_{1,1}\scriptstyle{(\frac{M}{2} + 1)}    \\
   \scriptstyle{\vdots}                \\
   INI_{1,1}\scriptstyle{(\frac{M}{2} + K - 1)} \\
\end{bmatrix} \hspace{-0.15 cm}=\hspace{-0.15 cm} 
}_{\mathbf{INI}_{1, 1}}
\underbrace{
\begin{bmatrix}
   1         &\hspace{-0.25 cm}\cdots &\hspace{-0.25 cm}{\omega}_{1}^{(\frac{M}{2})(M - 1)}       \\      
   1         &\hspace{-0.25 cm}\cdots &\hspace{-0.25 cm}{\omega}_{1}^{(\frac{M}{2} + 1)(M - 1)}   \\
   \vdots    &\hspace{-0.25 cm}\ddots &\hspace{-0.25 cm}\vdots                            \\
   1         &\hspace{-0.25 cm}\cdots &\hspace{-0.25 cm}{\omega}_{1}^{(\frac{M}{2} + K - 1)(M-1)} \\
\end{bmatrix}
\hspace{-0.1 cm}
\frac{1}{N}
\hspace{-0.1 cm}
\begin{bmatrix}
   {\bar\omega}_{0}^{(\frac{N}{2}-P)(N_{cp} + \frac{N}{2})}    &\hspace{-0.4 cm}\cdots  &\hspace{-0.2 cm}{\bar\omega}_{0}^{(\frac{N}{2}-2)(N_{cp} + \frac{N}{2})}     &\hspace{-0.30 cm}{\bar\omega}_{0}^{(\frac{N}{2}-1)(N_{cp} + \frac{N}{2})}     \\      
   \vdots                                   &\hspace{-0.4 cm}\ddots    &\hspace{-0.2 cm}\vdots  &\hspace{-0.30 cm}\vdots                                 \\
   {\bar\omega}_{0}^{(\frac{N}{2}-P)(N_{cp} + N - 1)} &\hspace{-0.4 cm}\cdots  &\hspace{-0.2 cm} {\bar\omega}_{0}^{(\frac{N}{2}-2)(N_{cp} + N - 1)} &\hspace{-0.30 cm}{\bar\omega}_{0}^{(\frac{N}{2}-1)(N_{cp} + N - 1)} \\
\end{bmatrix}
}_{\mathbf{W}_{1,1}^{INI}}
\hspace{-0.2 cm}
\underbrace{
\begin{bmatrix}
   X_{0}\scriptstyle{(\frac{N}{2}-P)} \\
   \scriptstyle{\vdots}        \\
   X_{0}\scriptstyle{(\frac{N}{2}-2)} \\
   X_{0}\scriptstyle{(\frac{N}{2}-1)} \\
\end{bmatrix}
}_{\mathbf{X}_{0}}
\tag{14}
\end{gather}
\vspace{-0.3 cm}
\begin{gather}
\underbrace{
\begin{bmatrix}
   INI_{0,0}\scriptstyle{(\frac{N}{2}-P)} \\
   \scriptstyle{\vdots}            \\
   INI_{0,0}\scriptstyle{(\frac{N}{2}-2)}  \\
   INI_{0,0}\scriptstyle{(\frac{N}{2}-1)}  \\
\end{bmatrix} \hspace{-0.15 cm}=\hspace{-0.15 cm}  
}_{\mathbf{INI}_{0, 0}}
\underbrace{
\begin{bmatrix}
   1         &\hspace{-0.25 cm}\cdots &\hspace{-0.25 cm}{\omega}_{0}^{(\frac{N}{2}-P)(\frac{N_{cp}}{2} + \frac{N}{2} - 1)}       \\      
   \vdots    &\hspace{-0.25 cm}\ddots &\hspace{-0.25 cm}\vdots                            \\
   1         &\hspace{-0.25 cm}\cdots &\hspace{-0.25 cm}{\omega}_{0}^{(\frac{N}{2} - 2)(\frac{N_{cp}}{2} + \frac{N}{2} - 1)}   \\
   1         &\hspace{-0.25 cm}\cdots &\hspace{-0.25 cm}{\omega}_{0}^{(\frac{N}{2} - 1)(\frac{N_{cp}}{2} + \frac{N}{2} - 1))} \\
\end{bmatrix}
\hspace{-0.1 cm}
\frac{1}{M}
\hspace{-0.1 cm}
\begin{bmatrix}{\scriptscriptstyle}
   {\bar\omega}_{1}^{(\frac{M}{2})(2M_{cp})}    &\hspace{-0.4 cm}{\bar\omega}_{1}^{(\frac{M}{2}+1)(2M_{cp})} &\hspace{-0.4 cm}\cdots &\hspace{-0.4 cm} {\bar\omega}_{1}^{(\frac{M}{2}+K-1)(2M_{cp})}   \\     
   {\bar\omega}_{1}^{(\frac{M}{2})(2M_{cp}+1)}  &\hspace{-0.4 cm}{\bar\omega}_{1}^{(\frac{M}{2}+1)(2M_{cp}+1)}  &\hspace{-0.4 cm}\cdots    &\hspace{-0.4 cm} {\bar\omega}_{1}^{(\frac{M}{2}+K-1)(2M_{cp}+1)} \\  
   \vdots                               &\hspace{-0.4 cm}\ddots                                 &\hspace{-0.4 cm}\vdots    &\hspace{-0.4 cm} \vdots                                  \\
   {\bar\omega}_{1}^{(\frac{M}{2})(M + M_{cp} -1)} &\hspace{-0.4 cm}{\bar\omega}_{1}^{(\frac{M}{2}+1)(M + M_{cp} -1)} & \hspace{-0.4 cm}\cdots    &\hspace{-0.4 cm} {\bar\omega}_{1}^{(\frac{M}{2}+K-1)(M + M_{cp} -1)} \\
\end{bmatrix}
}_{\mathbf{W}_{0,0}^{INI}}
\hspace{-0.2 cm}
\underbrace{
\begin{bmatrix}
   X_{1,0}\scriptstyle{ (\frac{M}{2}) }    \\
   X_{1,0}\scriptstyle{ (\frac{M}{2}+1) }   \\
   \scriptstyle{\vdots}           \\
   X_{1,0}\scriptstyle{ (\frac{M}{2}+K-1)} \\
\end{bmatrix}
}_{\mathbf{X}_{1,0}}
\tag{16}
\end{gather}
\begin{gather}
\underbrace{
\begin{bmatrix}
   INI_{0,1}\scriptstyle{(\frac{N}{2}-P)}  \\
   \scriptstyle{\vdots}            \\
   INI_{0,1}\scriptstyle{(\frac{N}{2}-2)}  \\
   INI_{0,1}\scriptstyle{(\frac{N}{2}-1)} \\
\end{bmatrix} \hspace{-0.15 cm} = \hspace{-0.15 cm} 
}_{\mathbf{INI}_{0, 1}}
\underbrace{
\begin{bmatrix}{\scriptscriptstyle}
   {\omega}_{0}^{(\frac{N}{2}-P)(\frac{N}{2} + \frac{N_{cp}}{2})}         &\hspace{-0.25 cm} \cdots &\hspace{-0.25 cm} {\omega}_{0}^{(\frac{N}{2}-P)(N - 1)}       \\      
   \vdots    &\hspace{-0.25 cm} \ddots &\hspace{-0.25 cm} \vdots                            \\
   {\omega}_{0}^{(\frac{N}{2}-2)(\frac{N}{2} + \frac{N_{cp}}{2})}         &\hspace{-0.25 cm} \cdots &\hspace{-0.25 cm} {\omega}_{0}^{(\frac{N}{2}-2)(N-1)}   \\
   {\omega}_{0}^{(\frac{N}{2}-1)(\frac{N}{2} + \frac{N_{cp}}{2})}        &\hspace{-0.25 cm} \cdots & \hspace{-0.25 cm}{\omega}_{0}^{(\frac{N}{2}-1)(N-1))} \\
\end{bmatrix}
\hspace{-0.1 cm}
\frac{1}{M}
\hspace{-0.1 cm}
\begin{bmatrix}{\scriptscriptstyle}
   {\bar\omega}_{1}^{(\frac{M}{2})(M - M_{cp})}     &\hspace{-0.25 cm} \cdots    &\hspace{-0.2 cm} {\bar\omega}_{1}^{(\frac{M}{2}+K-1)(M - M_{cp})}   \\    
      \vdots                                &\hspace{-0.25 cm} \ddots    & \vdots                                  \\
   1   &\hspace{-0.25 cm} \cdots    &\hspace{-0.2 cm} 1 \\  
   \vdots                                &\hspace{-0.25 cm} \ddots    & \vdots                                  \\
   {\bar\omega}_{1}^{(\frac{M}{2})(M-1)}  &\hspace{-0.25 cm} \cdots    &\hspace{-0.2 cm} {\bar\omega}_{1}^{(\frac{M}{2}+K-1)(M-1)} \\
\end{bmatrix}
}_{\mathbf{W}_{0,1}^{INI}}
\hspace{-0.2 cm} 
\underbrace{
\begin{bmatrix}
   X_{1,1}\scriptstyle{ (\frac{M}{2}) }    \\
   X_{1,1}\scriptstyle{ (\frac{M}{2}+1) }  \\
   \scriptstyle{\vdots}            \\
   X_{1,1}\scriptstyle{ (\frac{M}{2}+K-1)} \\
\end{bmatrix}
}_{\mathbf{X}_{1,1}}
\tag{18}
\end{gather}
\end{strip}

\subsection{INI from numerology 1-0 and 1-1 to numerology 0}

\begin{figure}[ht!] 
 \centering
 \includegraphics[width=0.45\textwidth] {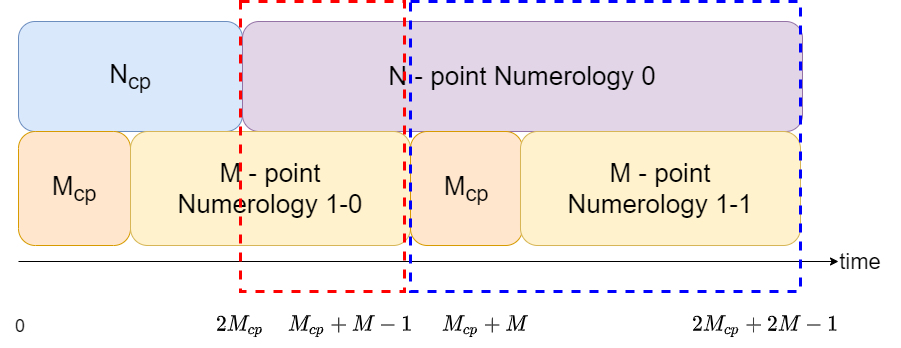}
 \caption{INI from numerology 1-0 and 1-1 to numerology 0 on time domain}
\end{figure}

Part of numerology 1-0, shown in red dashed lines in Fig. 2, between time samples $2M_{cp}$ and $M_{cp} + M - 1$ of subcarriers from $\frac{M}{2}$ to $\frac{M}{2} + K - 1$ causes INI on subcarriers from $\frac{N}{2}-P$ to $\frac{N}{2}-1$ of numerology 0. This expression can be modeled as a matrix form in equation (16), which can be simplified as

\begin{equation}
    {\mathbf{INI}_{0,0}} = {\mathbf{W}_{0,0}^{INI}} {\mathbf{X}_{1,0}}
\tag{17},
\end{equation}where,  $\textbf{\emph{INI}}_{0, 0}$ is a $P\times 1$ column vector that represents the INI created by numerology 1-0 on the last $P$ subcarriers of numerology 0. $\textbf{\emph{X}}_{1,0}$ is the $K\times 1$ column vector of symbols to be transmitted of numerology 1-0 and $\textbf{\emph{W}}_{0,0}^{INI}$ is the $P\times K$ matrix describing how the INI is created.

The whole part of numerology 1-1, shown in blue dashed lines in Fig. 2, between time samples $M_{cp} + M$ and $2M_{cp} + 2M - 1$ of subcarriers from $\frac{M}{2}$ to $\frac{M}{2} + K - 1$ causes INI on subcarriers from $\frac{N}{2}-P$ to $\frac{N}{2}-1$ of numerology 0. This expression can be modeled as a matrix form in equation (18), which can be simplified as

\begin{equation}
    {\mathbf{INI}_{0,1}} = {\mathbf{W}_{0,1}^{INI}} {\mathbf{X}_{1,1}}
\tag{19}
\end{equation}

We should combine the equations (17) and (19) to get the whole INI on numerology 0.

\begin{equation}
    {\mathbf{INI}_{0}} = {\mathbf{W}_{0,0}^{INI}}\ {\mathbf{X}_{1,0}} + {\mathbf{W}_{0,1}^{INI}} {\mathbf{X}_{1,1}}
    \tag{20}
\end{equation}

\subsection{INI Modelling and INI Pre-Equalization Matrix}

From equations (13), (15), (17), and (19), $K\times P$ matrices $\textbf{\emph{W}}_{1,0}^{INI}$, $\textbf{\emph{W}}_{1,1}^{INI}$  and $P \times K$ matrices $\textbf{\emph{W}}_{0,0}^{INI}$, $\textbf{\emph{W}}_{0,1}^{INI}$  are derived, respectively. By using equations (13), (15), and (20), the transmission symbols exposed to INI for each numerology can be written as

\begin{equation}
\begin{aligned}
    &\mathbf{X}_{0}^{INI} = \mathbf{X}_{0}  + \mathbf{W}_{0,0}^{INI} \mathbf{X}_{1,0} + \mathbf{W}_{0,1}^{INI} \mathbf{X}_{1,1}
    \\
    &\mathbf{X}_{1,0}^{INI} = \mathbf{X}_{1,0} + \mathbf{W}_{1,0}^{INI} \mathbf{X}_{0}
    \\
    &\mathbf{X}_{1,1}^{INI} = \mathbf{X}_{1,1} + \mathbf{W}_{1,1}^{INI}  \mathbf{X}_{0}
\end{aligned}
\tag{21}
\end{equation}


Equations in (21) can be shaped into a matrix form as
\begin{equation}
\begin{bmatrix}
  {\mathbf{X}_{0}^{INI}}    \\ \\
  {\mathbf{X}_{1,0}^{INI}}  \\ \\
  {\mathbf{X}_{1,1}^{INI}}  \\ 
\end{bmatrix} =
\begin{bmatrix}
 {\mathit{\mathbf{I_{P}}}}    & {\mathbf{W}_{0,0}^{INI}} & {\mathbf{W}_{0,1}^{INI}} \\ \\
 {\mathbf{W}_{1,0}^{INI}} & {\mathit{\mathbf{I_{K}}}}    & {\mathit{\mathbf{0_{K}}}}    \\ \\
 {\mathbf{W}_{1,1}^{INI}} & {\mathit{\mathbf{0_{K}}}}    & {\mathit{\mathbf{I_{K}}}}    \\
\end{bmatrix}
\begin{bmatrix}
  {\mathbf{X}_{0}}   \\ \\
  {\mathbf{X}_{1,0}} \\ \\
  {\mathbf{X}_{1,1}} \\

\end{bmatrix} 
\tag{22}
\end{equation}where ${\textbf{\emph{I}}_{P}}$ and ${\textbf{\emph{I}}_{K}}$ denote $P\times P$ and $K\times K$ size identity matrices, respectively. ${\textbf{\emph{O}}_{K}}$ denotes the $K\times K$ size zero matrix. Therefore, equation (22) can be written as
\begin{equation}
    {\mathbf{X}^{INI}} = {\mathbf{W}^{INI}}  {\mathbf{X}}
    \tag{23}
\end{equation}where ${\textbf{\emph{X}}^{INI}}$ is a $(P+2K)\times 1$ column vector that denotes transmission symbols exposed to INI. ${\textbf{\emph{W}}^{INI}}$ is the $(P+2K)\times(P+2K)$ matrix that models how INI is created at each numerologies' transmission symbols and ${\textbf{\emph{X}}}$ is the $(P+2K)\times 1$ column vector of transmission symbols that are exposed to INI. In other words, INI-free transmission symbols. Therefore, equation (23) models how transmission symbols are to be exposed to INI that occurs in the generation of the multi-numerology signals at the transmitter.

To remove INI on the transmission subcarriers (i.e symbols) completely, transmission symbol vector, $\mathbf{X}$, should be pre-equalized. To do that, the inverse of ${\textbf{\emph{W}}^{INI}}$ should be used.

\begin{equation}
    {\mathbf{X}^{b}} = {(\mathbf{W}^{INI})^{-1}} {\mathbf{X}}
    \tag{24}
\end{equation}Here, $(\mathbf{W}^{INI})^{-1}$ denotes the pre-equalization matrix, and $\textbf{\emph{X}}^{b}$ becomes a pre-equalized transmission symbol vector. In multi-numerology signal creation, using $\textbf{\emph{X}}^{b}$ instead of $\textbf{\emph{X}}$ will remove the INI on $\textbf{\emph{X}}$ as 
\begin{equation}
    {\mathbf{X}} = {(\mathbf{W}^{INI})} {\mathbf{X}^{b}}
    \tag{25}
\end{equation}
Briefly, in order to eliminate INI on transmission symbols in multi-numerology creation, transmission symbols need to be pre-equalized by ${(\mathbf{W}^{INI})^{-1}}$. Note that, this is a fixed matrix and does not depend on instantaneously transmitted data. Therefore it is computed only once and used for all OFDM symbols.  

\section{Numerical Results and Discussion}
In this section, the performance of the proposed INI pre-equalization technique is evaluated in additive white Gaussian noise added Rayleigh fading channels via Monte Carlo simulations. In the simulations, we consider $P = 96$ and $K = 48$ to give equal bandwidth for each numerology and only these bands will be considered for INI pre-equalization.
Channel coding introduced in 5G and beyond is not considered throughout the simulations because the aim is to remove the INI that occurred on the transmission symbols.
The total frequency bandwidth is divided equally for numerology 0 and 1. For this scenario, $256$ and $128$-point IDFT/DFT sizes are assigned to numerology $0$ and $1$, respectively. The guard band is not used between the numerologies and numerologies' CP lengths are chosen as $18$ and $9$, respectively. We consider 64-QAM and 256-QAM modulations for simulations.

First, generated multi-numerology OFDM signals via INI pre-equalization are directly demodulated without any noise and channel effect. After demodulation, it has been observed that the EVM values of the transmission symbols are 0 dB.

\begin{figure} [ht!] 
 \centering
 \includegraphics[width=0.40\textwidth]{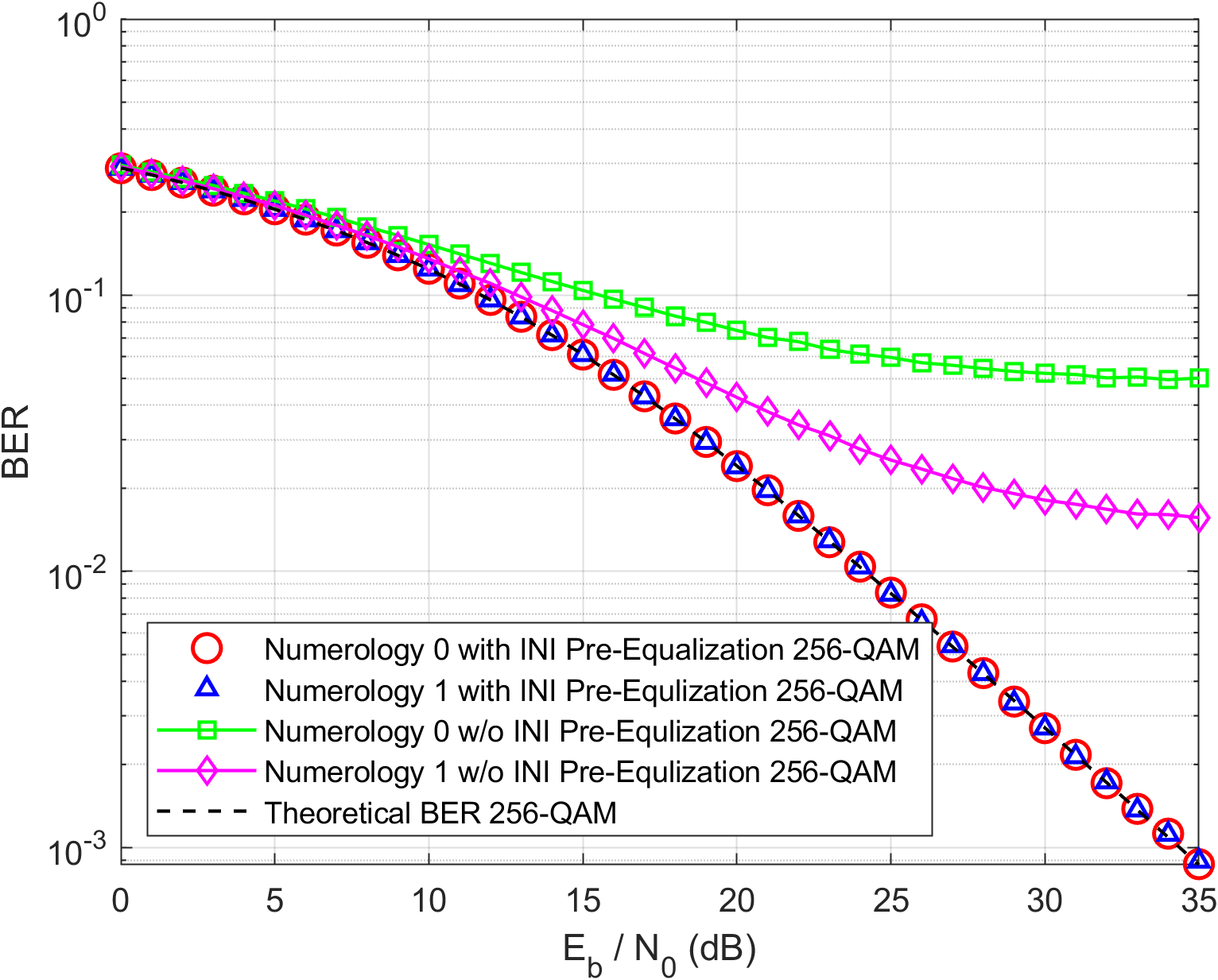}
 \caption{BER results of 256-QAM for numerology 0 and 1}
\end{figure}

\begin{figure} [ht!] 
 \centering
 \includegraphics[width=0.40\textwidth]{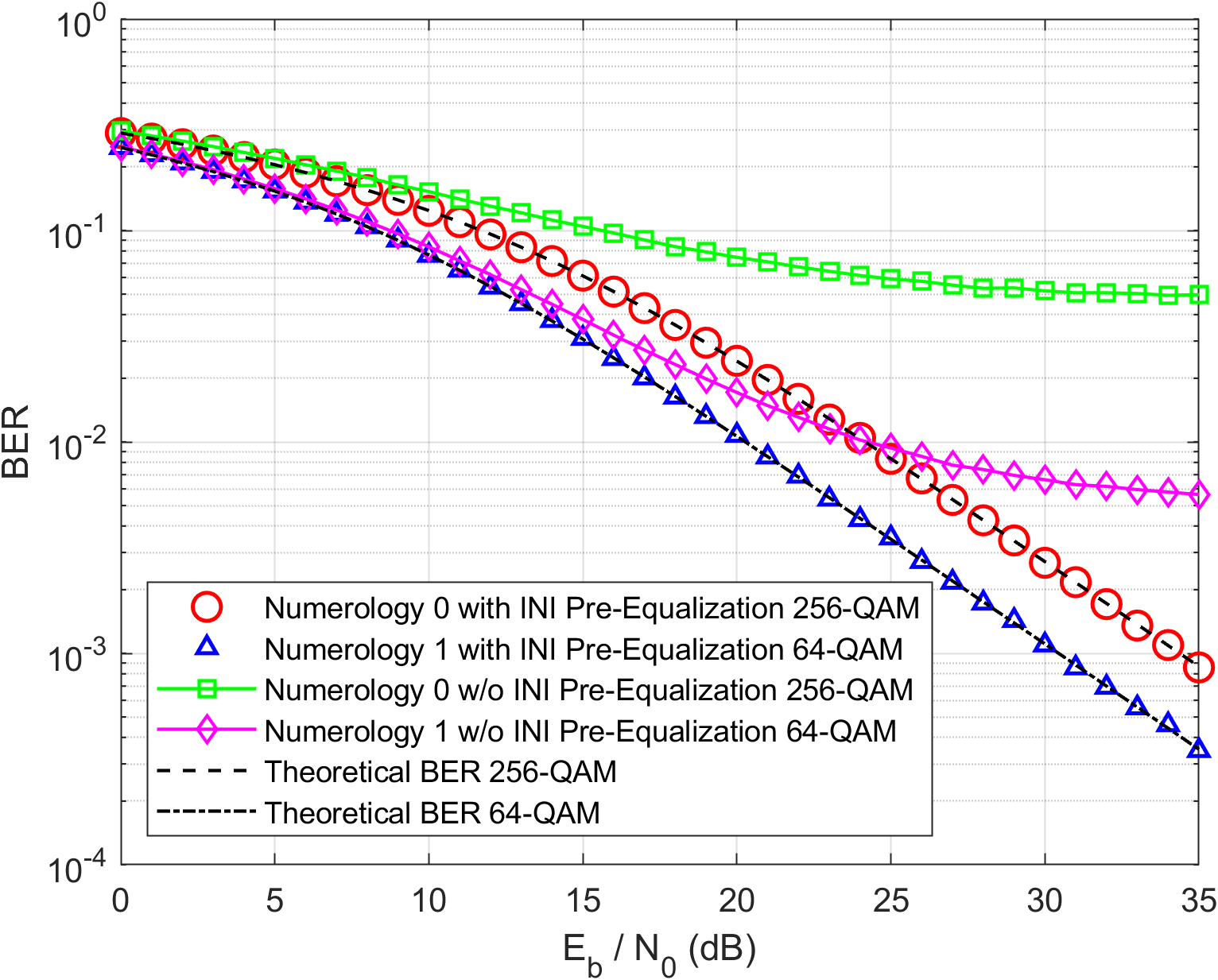}
 \caption{BER results of 256-QAM and 64-QAM for numerology 0 and 1}
\end{figure}

In Fig. 3, and 4, performances of INI pre-equalizer techniques are depicted by circle and triangle markers. Theoretical BER performance (assuming no INI) is considered as the benchmark and the cases where the INI is not pre-equalized are shown with square and diamond markers. In Fig. 3 and 4, the BER performance of both numerologies are aligned with the theoretical BER in Rayleigh fading channel for both 64-QAM and 256-QAM. 

We didn't compare our scheme with the guard-band-based schemes, since those methods result in a significant decrease in throughput. Besides, the performance of pre-equalizer achieves the theoretical BER limits, which is already the best possible performance.

\begin{figure} [ht!] 
 \centering
 \includegraphics[width=0.40\textwidth]{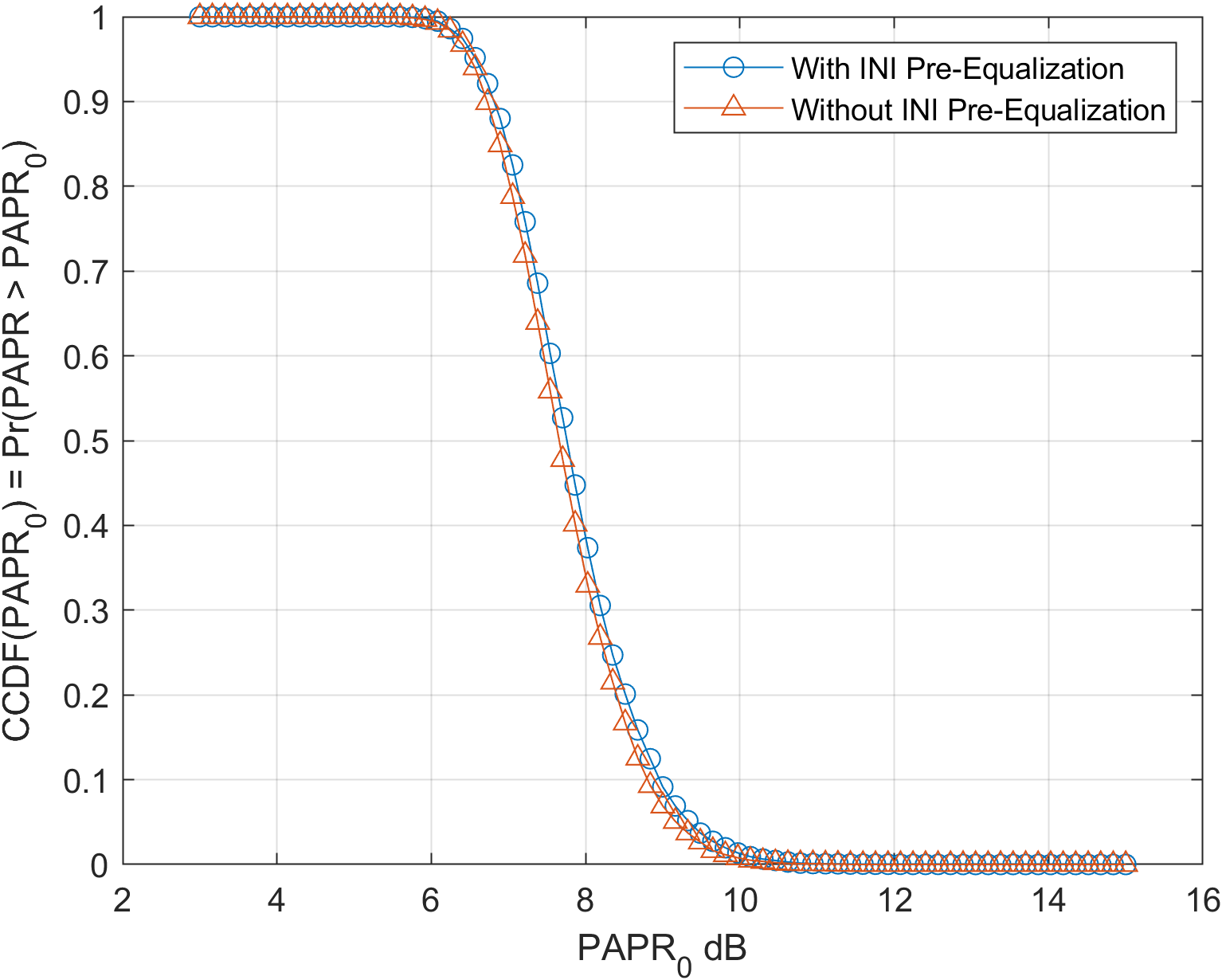}
 \caption{PAPR CCDF results of 256-QAM for numerology 0 and 1}
\end{figure}

In Fig. 5, the PAPR analysis of INI pre-equalizer is investigated. Complementary cumulative distribution functions (CCDF) is taken into account. CCDF PAPR of pre-equalized multi-numerology signal is depicted by a circle and CCDF PAPR of non-pre-equalized multi-numerology signal is depicted by a triangle markers. The pre-equalization technique makes almost no changes to the PAPR while removing the INI.

Please note that only one ${(\textbf{\emph{W}}^{INI})^{-1}}$ square matrix of size $(P+2K)\times (P+2K)$ from (24) is used as a pre-equalizer to remove INI for any modulation scheme. This matrix only depends on the IDFT/DFT sizes, CP length, adjacent $P$ and $K$ subcarriers, and chosen numerology indices. After the pre-equalizer matrix is created once for a system, then the same matrix can be used for any modulations. Similar INI pre-equalizer matrices could be derived for other numerology pairs as $\mu = 0, 2$ and $\mu = 1, 2$ for any $P$ and $K$ values.

\section{Conclusion}
In this work, a general solution for the removal of Inter numerology Interference (INI) in multi-numerology OFDM systems is proposed. The INI is mathematically expressed as a weighted linear combination of the transmitted symbols and the INI creation matrix is derived. Then, using an inverse of the INI creation matrix as a pre-equalizer we can completely remove the INI on the transmitter side for both numerologies. As future work, it is possible to further investigate to implement the proposed technique on MIMO systems and windowed-OFDM signals.
\ifCLASSOPTIONcaptionsoff
  \newpage
\fi



\bibliographystyle{IEEEtran}
\bibliography{references}
%

%








\end{document}